\pdfoutput=1
\documentclass[journal=jacsat,manuscript=article]{achemso}

\usepackage{caption}
\usepackage{subcaption}

\usepackage{adjustbox}
\usepackage{booktabs}
\usepackage{amssymb}
\usepackage{bm}
\usepackage{xfrac}
\usepackage{pifont}
\mciteErrorOnUnknownfalse

\usepackage{algorithm}
\usepackage{algcompatible}

\usepackage{chemformula} 
\usepackage[T1]{fontenc} 

\author{Anna C. M. Thöni}  
\email{chiara.thoeni@ru.nl}\affiliation[Radboud University]{Donders Centre for Cognition, Radboud University, Nijmegen, the Netherlands}
\author{William E. Robinson}\affiliation[Radboud University]{Institute for Molecules and Materials, Radboud University, Nijmegen, the Netherlands}
\author{Yoram Bachrach}\affiliation{Meta FAIR, London, United Kingdom}
\author{Wilhelm T. S. Huck}\affiliation[Radboud University]{Institute for Molecules and Materials, Radboud University, Nijmegen, the Netherlands}
\author{Tal Kachman}\affiliation[Radboud University]{Donders Centre for Cognition, Radboud University, Nijmegen, the Netherlands}

\title[Modelling Chemical Reaction Networks using Neural Ordinary Differential Equations]
  {Modelling Chemical Reaction Networks using Neural Ordinary Differential Equations}

\begin{document}

\begin{abstract} 
In chemical reaction network theory, ordinary differential equations are used to model the temporal change of chemical species concentration. As the functional form of these ordinary differential equations systems is derived from an empirical model of the reaction network, it may be incomplete. Our approach aims to elucidate these hidden insights in the reaction network by combining dynamic modelling with deep learning in the form of neural ordinary differential equations. Our contributions not only help to identify the shortcomings of existing empirical models but also assist the design of future reaction networks. 

\end{abstract}

\section{Introduction}
 The dynamic behaviour of networks of chemical reactions is typically described using a system of ordinary differential equations (ODEs). Such systems of ODEs are derived by combining the chemical reaction network (CRN), which underlies the observed dynamics, with the law of mass action \cite{MassAction, erdi1989mathematical, kachman2017self}. The parameters of the ODEs, such as the rate constant of a reaction or the concentration of a reactant, are ideally experimentally determined or need to be calibrated to maximize the similarity between the model predictions and the experimental data over time \cite{ODEcalibration, lakrisenko2023efficient}. 

Despite this calibration, it is not guaranteed that the ODEs can predict the experimental data perfectly, as the data can be noisy or incomplete \cite{jaqamanlinking}. The proposition of mass action assumes that the reaction mixture is well-mixed and under thermal equilibrium but these requirements are not always fulfilled in practice.  Furthermore, the differential equations cannot describe any dynamics that are not included in the theoretical specification of the CRN. This limits the modelling performance when the CRN is misspecified, for example by the presence of hidden interactions or competition between the species involved \cite{rondelez2012competition, wen2023chemical}. The limited theoretical modelling performance may hamper the quality of the predicted behaviour of chemical systems consisting of multiple reactions. 

We propose to improve the predictive behaviour of ODEs by combining dynamic systems modelling with deep learning in the form of Neural ODEs (nODEs)~\cite{chen2018neural}. nODEs are akin to deep learning models~\cite{lecun2015deep} which drive many recent successes, from self-driving cars~\cite{badue2021self}, through large language models~\cite{brown2020language, Hodl2023} to protein structure prediction~\cite{jumper2021highly}. Deep learning models use artificial neural networks, mathematical models that can be trained to approximate any function \cite{HORNIK1989359}. However, nODEs are inspired by the advances in deep residual learning and normalizing flows, which build complex transformations by applying a sequence of smaller transformations \cite{he2015deep, dinh2016density}; unlike standard deep learning models that operate through a discrete sequence of hidden layers, nODEs employ a neural network to {\em parameterize the derivative of a hidden state}. As such, the nODE output is computed through a differential equation solver. Thus, in nODEs, neural networks represent the smaller transformations, and in the limit, the composition of small neural transformations behaves like a differential equation \cite{chen2018neural}. However, in contrast with a predefined differential equation, the nODE takes form during training \cite{kidger2020neural}, allowing the nODE to minimize the difference between the observed and modelled data. Therefore, we hypothesize that the predictions of the nODEs are of a better quality than those from the original theoretical model. 

The possibility of solving chemical kinetics using nODEs has been explored before. Nonetheless, the key limitations of these earlier methods are that they require that the number of reactions is known a priori \cite{ji2021autonomous} or focus on the thermodynamical state of the reactions \cite{ChemNode}. In contrast, we aim to address the differences between the modelled and experimental concentrations {\em without making any assumptions about the reactions within the CRN}. To this end, we follow \citet{rackauckas2020universal} and augment the existing theoretical system of ODEs with a neural network component. The neural network acts as a correction term: it captures the trends within the experimental data that the theoretical model does not explain. The network contribution can be calculated for each measured chemical species. The separate contributions provide insight into the magnitude and temporal signature of the differences between the theoretical model and the experimental data.

We test our methods by predicting the phase space of an oscillatory system. We focus on oscillating behaviour as the quality of the model prediction is easily jeopardized by hidden interactions between the species involved. Predicting the amplitude, periodicity, and phase space accurately is challenging as the model needs to account for noisy measurements, hidden interactions and dynamics that are out of equilibrium. Therefore, we split the prediction of the oscillator dynamics into smaller objectives. We first consider two experimental datasets collected by \citet{catalyticOscillator}, who composed a chemical oscillator of reactions between small organic molecules. The CRN consists of four reactions and involves 7 species. The reactions include the autocatalytic Fmoc-piperidine (2) deprotection via dibenzofulvene (6); the N-methylpiperidine-catalysed (5) Fmoc-piperidine deprotection; the fast inhibition via acetylation by \textit{p}-nitrophenyl acetate (3); and the slow inhibition by phenyl acetate (4) that converts piperidine (1) to N-acetyl piperidine (7) \cite{catalyticOscillator}. We use the data from the single-pulse (aperiodic, batch) experiment as well as the measured series of sustained oscillations, which require an open system (flow reactor). Furthermore, we test the model's performance on hidden interactions by deliberately leaving out reactions from the CRN described above. Finally, we predict the oscillation periodicity and space of a CRN by transferring the dynamics captured by the nODE between experiments.

The results found offer two insights. Firstly, the neural network contributions can be used to verify the correctness of the theoretical model under noisy measurements. Secondly, the contributions can be used to account for and uncover any dynamics that are not included in the theoretical model. Combining these findings, we show that nODEs are more accurate predictors of the period of oscillating concentrations of unseen experimental settings than the theoretical system of ODEs alone.

\section*{Methods}
\paragraph{Chemical Reaction Networks and nODEs}
Before introducing our nODE approach, we briefly discuss dynamical systems modelling with ODEs. As mentioned in the Results, the concentration of chemical species can be modelled by solving the IVP constrained by a system of ODEs. More formally, let the dynamical system be defined from time $t_0$ until $T$, where $y$ describes the unknown solution represented by the initial point $y_0$ and $h(t, y(t))$ a system of ODEs. With these terms, the IVP is defined in Equation \ref{eq: IVP}.
\begin{equation}
    y(t_0) = y_0 \;\;\;\;\;\;\;\; \frac{dy}{dt}(t) = h(t, y(t))
    \label{eq: IVP}
\end{equation}
While the solution to Equation \ref{eq: IVP} equals $y = \int_{t_0}^T h(t, y(t))dt$, systems of differential equations are often solved using numerical methods due to the unavailability of the antiderivative of $h$ \cite{Calculus2013}. Numerical methods calculate the full time-series $y$ step-by-step, starting at $y_0$. When CRNs are considered, $h$ represents the theoretical system of ODEs satisfying the law of mass action, $y$ the concentrations of the chemical species over time, $y_0$ the initial concentration and $\sfrac{dy}{dt}$ the change in concentration over time $t \in [t_0, T]$.

In nODEs, and universal differential equations in particular, the existing theoretical system of ODEs $h_\kappa$ is augmented with the data-driven neural network $f_\theta$ \cite{rackauckas2020universal}. The resulting system of ODEs is presented in Equation \ref{eq: UDE}.
\begin{equation}
    \frac{dy}{dt}(t + \delta t) = h_\kappa(t, y(t)) + f_\theta(t, y(t))
    \label{eq: UDE}
\end{equation}
Here, $\kappa$ represents the parameters of $h$, in our case the vector of reaction rate coefficients $k_{tr},\; k_{ac},\; k_{inh1}\text{ and }k_{inh2}$, retrieved from \citet{catalyticOscillator}, $h$ represents the full system of ODEs, describing the rate of change of each chemical species in the oscillating CRN (Equation \ref{eq: ODE}). The concentration of \textit{N}-methylpiperidine (5) stays constant over time and is therefore not included. The species are abbreviated with their index on the right-hand side of the equation.
\begin{align}
\begin{split}
    \frac{d[\text{piperidine (1)}]}{dt} &= k_{tr}[5][2] + k_{ac}[1][2] - k_{inh1}[1][3] \\
    &\phantom{=}- k_{inh2}[1][4] - sv[1]\\
    \frac{d[\text{Fmoc-piperidine (2)}]}{dt} &= -k_{tr}[5][2] - k_{ac}[1][2] + sv\left([2]_{in}-[2]\right)\\
    \frac{d[\text{\textit{p}-nitrophenyl acetate (3)}]}{dt} &= -k_{inh1}[1][3] + sv([3]_{in}-[2])\\
    \frac{d[\text{phenyl acetate (4)}]}{dt} &= -k_{inh2}[1][4] + sv([4]_{in}-[4])\\
    \frac{d[\text{dibenzofulvene (6)}]}{dt} &= k_{tr}[5][2] + k_{ac}[1][2] -sv[6]\\
    \frac{d[\text{\textit{N}-acetyl piperidine (7)}]}{dt} &= k_{inh1}[1][3] + k_{inh2}[1][4] - sv[7]
\end{split}
\label{eq: ODE}
\end{align}
The stepsize $dt$ is governed by the differential equation solver. We use Kværnø's $\sfrac{5}{4}$ method to account for the stiffness of the ODE \cite{kvaerno2004singly, kidger2020neural}. Furthermore, we change the temporal scale of the experimental data from seconds to hours (single-pulse) or days (oscillations) to reduce the total number of steps the solver makes. 
\paragraph{Neural Network Architecture}
The neural network used consists of a LSTM cell with a 32-dimensional hidden state followed by a linear layer of 32 neurons. The network uses identity activations before and after the application of the linear layer. The number of inputs and outputs of the network matches the number of species that have been measured during the experiments. At each step within the equation solve, the neural network is presented with the measurements at a single time point instead of the full time series. Still, the LSTM outperforms a fully connected neural network during the experiments with synthetic data (see the Supporting Information). The model has been implemented using JAX \citep[version 0.4.14]{deepmind2020jax} and Equinox \citep[version 0.11.1]{kidger2021equinox}.
\paragraph{Neural ODE Training and Inference} Using Equation \ref{eq: UDE} and the initial concentrations of each species, we train $f_\theta$ to match the experimental data at the measurement times $t$. The network parameters $\theta$ are updated according to a mean squared error (MSE) loss with respect to the species that have been measured. Our data consists of the mean and standard deviation calculated over two experimental measurements. We generate artificial training, validation and test data by drawing from a normal distribution that is parameterized with these statistics. The neural network has been trained on $N=1000$ train samples using an Adabelief optimizer with a learning rate of $6 \times 10^{-3}$ \cite{zhuang2020adabelief, deepmind2020jax}. We did not apply dropout as this was found to cause instabilities during the differential equation solve, which resulted in a large number of rejected solver steps. For the oscillating data, we first train the nODE on the first 10\% of each time series to prevent getting stuck in a local minimum \cite{kidger2020neural}. Next to the nODE predictions, we calculate the separate neural network contribution $\sfrac{dy_{f_\theta}}{dt}$ given the prediction of the complete differential equation $\sfrac{dy}{dt}$.
\begin{equation}
    \frac{{dy}_{f_\theta}}{dt}(t_{n+1}) = f_\theta(t_{n},y(t_{n}))
    \label{eq: nn contribution}
\end{equation}

\paragraph{The Identification of Missing Reactions}
We use a similar modelling approach for the identification of missing reactions. However, instead of subtracting a constant concentration for each time period, we remove a part of the theoretical system of ODEs. In particular, we modify the change in N-acetyl piperidine to the definition presented in Equation \ref{eq: missing reaction}. This modification partially removes the pathway of the second inhibition reaction of the enzymatic oscillator. The rest of the theoretical system of ODEs remains unaltered. 
\begin{equation}
    \frac{d[\text{\textit{N}-acetyl piperidine (7)}]}{dt} = k_{inh1}[1][3] - sv[7]
    \label{eq: missing reaction}
\end{equation}

\section{Results and Discussion}
\begin{figure*}[!b]
\centering
\begin{subfigure}[b]{0.49\textwidth}
         \centering
         \includegraphics[width=\textwidth]{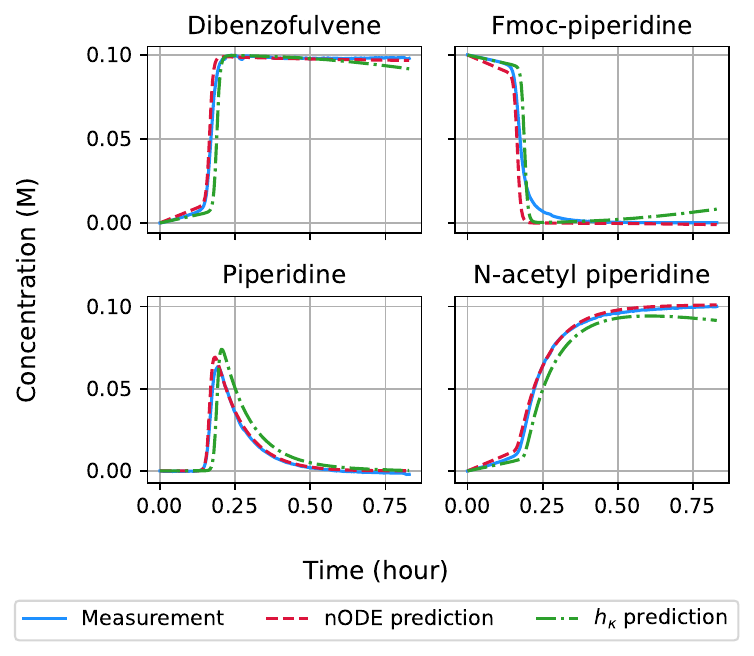}
          \caption{}
     \end{subfigure}
\hfill
\begin{subfigure}[b]{0.49\textwidth}
         \centering
         \includegraphics[width=\textwidth]{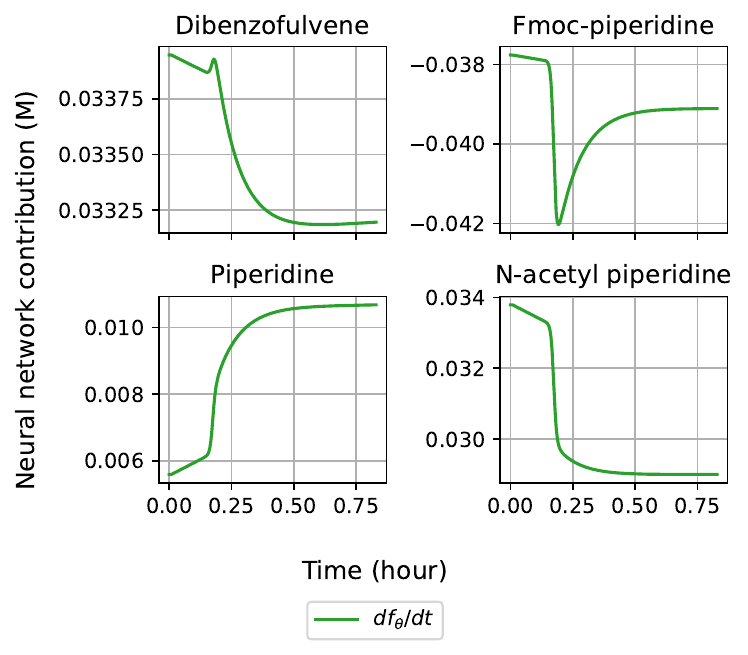}
          \caption{}
\end{subfigure}
\vspace{-3mm}
\caption{The predictive performance of the nODE on the single-pulse data. (a), the experimental measurements (solid blue), predictions from the nODE (dashed red) and $h_\kappa$ (dashdotted green). The four panes illustrate the molar concentrations for the species dibenzofulvene, Fmoc-piperidine, piperidine and N-acetyl piperidine. (b), the neural network contribution for the four measured species.} \label{fig: single_pulse_overview}
\end{figure*}
\subsection{Modelling Experimental Data using nODEs} We first consider the performance of the ``standard" nODE, where $h_\kappa$ is based on just the theoretical model. Figure \ref{fig: single_pulse_overview} illustrates the results for the single-pulse experiment based on this model. The experimental measurements (solid blue), nODE predictions (dashed red) and predictions based on $h_\kappa$ (dashdotted green) for the species Fmoc-piperidine, dibenzofulvene, piperidine and N-acetyl piperidine are shown in Figure \ref{fig: single_pulse_overview}a. The figure shows that the nODE provides a better fit than $h_\kappa$. This relative improvement can be attributed to the positive neural network contributions for the species dibenzofulvene and piperidine and N-acetyl piperidine, and the negative contribution for  Fmoc-piperidine (Figure \ref{fig: single_pulse_overview}b). The neural network contributions change over time. The rate of their change is the strongest at the start of the reaction and tends to flatten towards the end, indicating convergence towards a steady state. 
\subsubsection{Compensating for Hidden Interactions} 
\begin{figure}[H]
    \centering
    \begin{subfigure}[b]{0.35\textwidth}
         \centering
         \includegraphics[width=\textwidth]{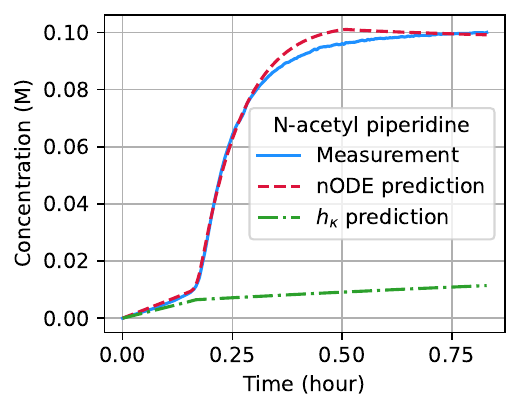}
          \caption{}
\end{subfigure}
\begin{subfigure}[b]{0.35\textwidth}
         \centering
         \includegraphics[width=\textwidth]{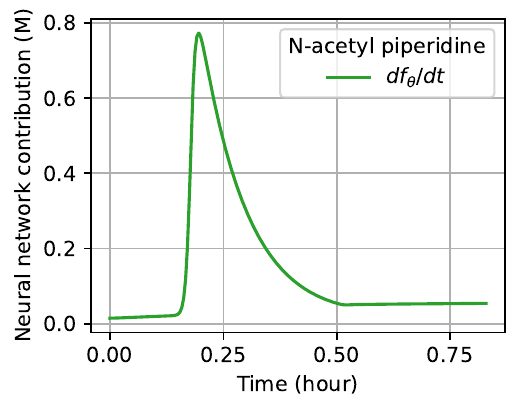}
         \caption{}
\end{subfigure}
\vspace{-3mm}
\caption{The predictive performance of the nODE for which the reaction for N-acetyl piperidine is hidden. (a), the experimental measurements (solid blue), predictions from the nODE (dashed red) and $h_\kappa$ (dashdotted green) for N-acetyl piperidine. (b), the neural network contribution.}
\label{fig: missing_reactions_PipAct}
\end{figure}

We demonstrate the resilience of the nODE against hidden interactions that are not included in the theoretical model of the CRN. To this end, we adapt the theoretical model $h_\kappa$ by removing the slow inhibition pathway via phenyl acetate, which negatively affects the creation of N-acetyl piperidine. We assess the modelling performance under this condition by focussing on the concentrations of N-acetyl piperidine, shown in Figure \ref{fig: missing_reactions_PipAct}. Even though the nODE does not provide a perfect fit (Figure \ref{fig: missing_reactions_PipAct}a), the missing variable is largely accounted for by an increase in the neural network contribution in Figure \ref{fig: missing_reactions_PipAct}b compared to Figure \ref{fig: single_pulse_overview}b. The predicted concentrations of the remaining species can be found in the Supporting Information.

\subsubsection{Oscillating Reaction Networks}
\begin{figure}[H]
    \centering
    \begin{subfigure}[b]{0.8\textwidth}
         \centering
         \includegraphics[width=\textwidth]{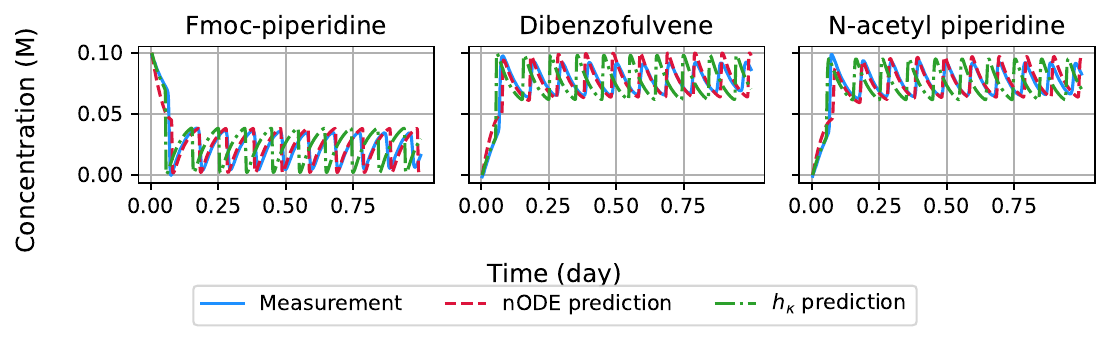}
         \caption{}
\end{subfigure}
\hfill
    \begin{subfigure}[b]{0.8\textwidth}
         \centering
         \includegraphics[width=\textwidth]{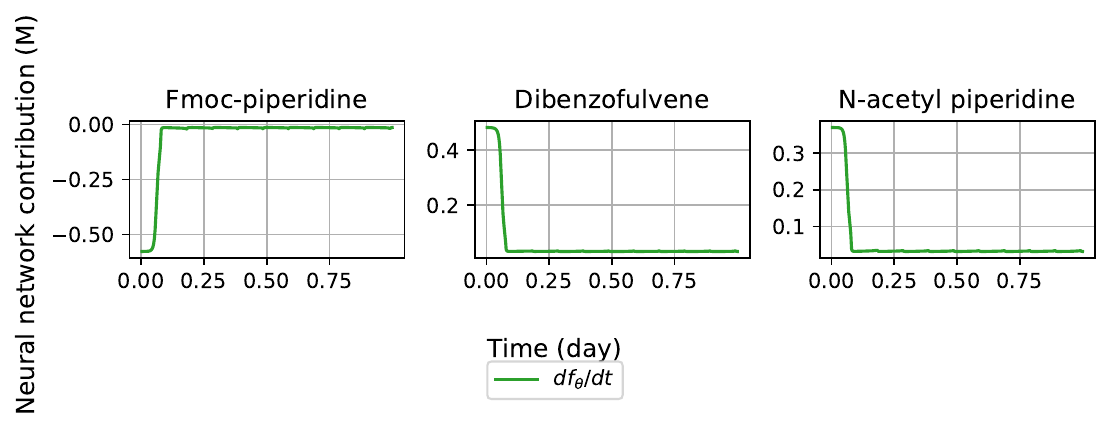}
         \caption{}
\end{subfigure}
\vspace{-3mm}
\caption{The predictive performance of the nODE on the oscillating open system for dibenzofulvene. (a), the experimental measurements (solid blue), predictions from the nODE (dashed red) and $h_\kappa$ (dashdotted green). (b), the neural network contribution.}
\label{fig: os}
\end{figure}

Lastly, we evaluate the predictive performance of the nODE on the oscillating CRN. The results of this experiment are shown in Figure \ref{fig: os}. When comparing the predictions of the nODE and $h_\kappa$ in Figure \ref{fig: os}a, it becomes clear that the oscillations within the data have a different frequency than those predicted by $h_\kappa$. The neural network contribution (Figure \ref{fig: os}b) ensures that the oscillations predicted by the nODE are more time-locked with the data. This effect is particularly strong at the start of the reaction as the data and the predictions tend to go out of phase later. As the observed phase changes over time, the improved time-keeping ability of the nODE compared to $h_\kappa$ demonstrates that these changes can be approximated with a neural network.

\subsubsection{Interpreting the Neural Network Contribution}
As discussed above, the separate neural network contributions provide insight into the magnitude and temporal signature of the discrepancies between the theoretical ODEs and the data. These insights may be translated into concrete changes to the theoretical model with the application of symbolic regression \cite{symbolicRegression}. Still, the neural network contributions may be hard to interpret, as they aggregate over all forms of noise within the system. A different use case of the nODEs focuses on transferring the learned dynamics between experiments. Doing so, we assume that the noise around the observations is independent and identically distributed (iid.) across experimental settings. While this may be a strong assumption, it is partially fulfilled as long as the experiments are carried out under the same conditions. Transferring information between experimental settings can be beneficial when predicting experimental outcomes. The next section explores this idea in the context of oscillating CRNs. 

\subsection{Predicting the Oscillation Space}
In oscillating CRNs, sustained oscillations are only observed for a relatively narrow set of input values for the control parameters. Outside this regime, the system can either show a steady state or damped oscillations that decay towards a steady state. The concentrations of species flowing into the reactor must therefore be chosen carefully. This is a major experimental challenge, as the system of ODEs often does not predict the oscillatory regime quantitatively correct, and experiments are laborious and time-consuming. We use the nODE to predict the regime of sustained oscillations for various concentrations of Fmoc-piperidine and phenyl acetate that are supplied in flow. As these compounds are part of the positive and negative feedback loops within the CRN, any variation of their input values affects the phase and stability of the oscillations. As a consequence, the oscillation space of the experiment is described by the Cartesian product of the different concentrations in flow. It is important to note that we do not train a nODE for each experimental setting. Instead, we train nODE once and transfer the learned dynamics between experiments. 

\begin{figure}
    \centering
    \begin{subfigure}[b]{0.32\textwidth}
         \centering
         \includegraphics[width=\textwidth]{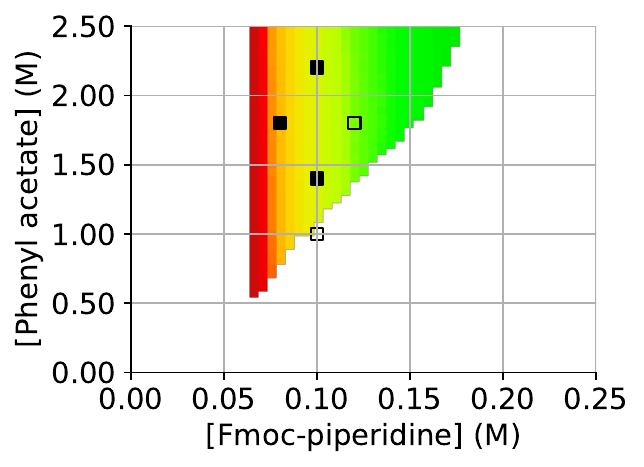}
         \caption{}
    \end{subfigure}
    \hfill
        \begin{subfigure}[b]{0.32\textwidth}
         \centering
         \includegraphics[width=\textwidth]{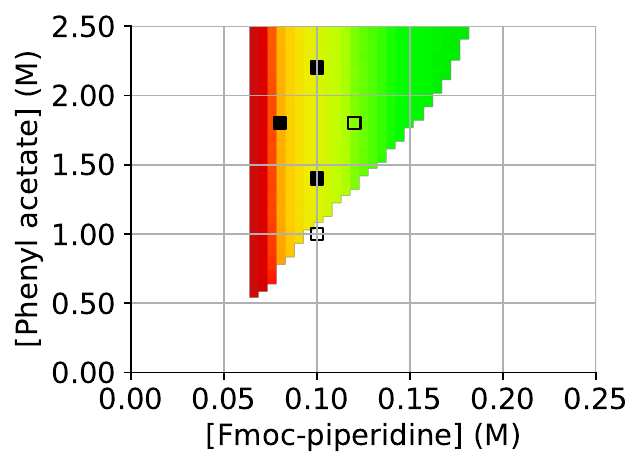}
         \caption{}
    \end{subfigure}
    \hfill
        \begin{subfigure}[b]{0.32\textwidth}
         \centering
         \includegraphics[width=\textwidth]{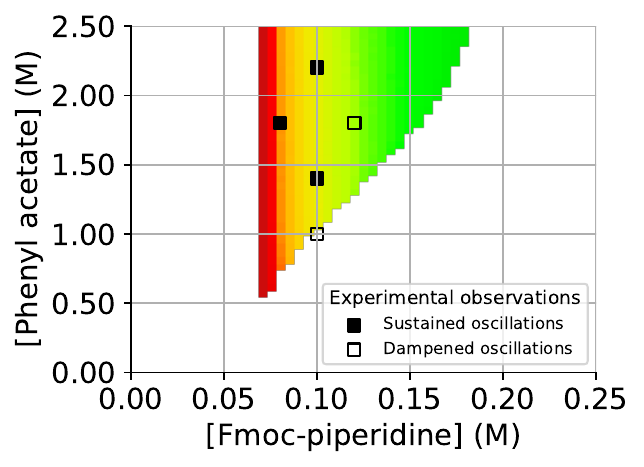}
         \caption{}
    \end{subfigure}
        \begin{subfigure}[b]{0.2\textwidth}
         \centering
         \includegraphics[width=\textwidth]{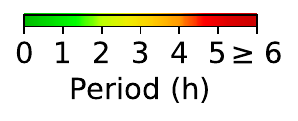}
    \end{subfigure}
    \caption{The oscillation spaces predicted by (a) the theoretical model, (b) the nODE trained on the complete time series and (c) the nODE trained on the first half of the time series. The coloured regions show the predicted period of the oscillations. Outside of these regions, the models predict damped or no oscillations. The filled squares indicate experimentally observed sustained oscillations, the open squares are associated with damped oscillations \cite{catalyticOscillator}.}
    \label{fig: phase diagrams}
\end{figure}

The predictions of the theoretical model, a nODE trained on the first half of the oscillations and a nODE trained on all oscillations are shown in Figure \ref{fig: phase diagrams}. The models are trained on data from the experimental setting where 100mM Fmoc-piperidine and 1.8 M phenyl acetate are presented in flow. Once the models are trained, they are presented with different concentrations of phenyl acetate and Fmoc piperidine in flow, spanning the oscillation space. The predicted oscillation spaces are compared to experimental observations, indicated with (filled) squares. The area in which sustained oscillations are predicted does not differ greatly between the models: all models make a correct prediction for four out of five observed oscillations. 

The fifth experiment, where 120mM Fmoc-piperidine and 1.8M phenyl acetate are supplied in flow, is incorrectly predicted as a source of stable oscillations. Even after having trained on the data collected during all five experiments, the nODE was not able to correctly adjust the regime of stable oscillations. This limitation could be fueled by a variable initialization time between experimental setups, violating the assumption that the noise is iid. between experiments. For example, it is possible that the chemicals are not instantaneously well-mixed in the continuous stirred-tank reactor (CSTR), which is an important assumption underlying all work in such reactors. Breaking this fundamental assumption would introduce a large discrepancy between the theoretical model and the measured data. If the initialization time varies between experiments, then the nODE may learn spurious relationships that inhibit the generalization between stable and damped oscillations. When the latter is combined with the small number of experiments under consideration, it can be very hard to learn transitions between stable and decaying oscillations. 

\begin{figure}[H]
    \centering
    \begin{subfigure}[b]{0.32\textwidth}
         \centering
         \includegraphics[width=\textwidth]{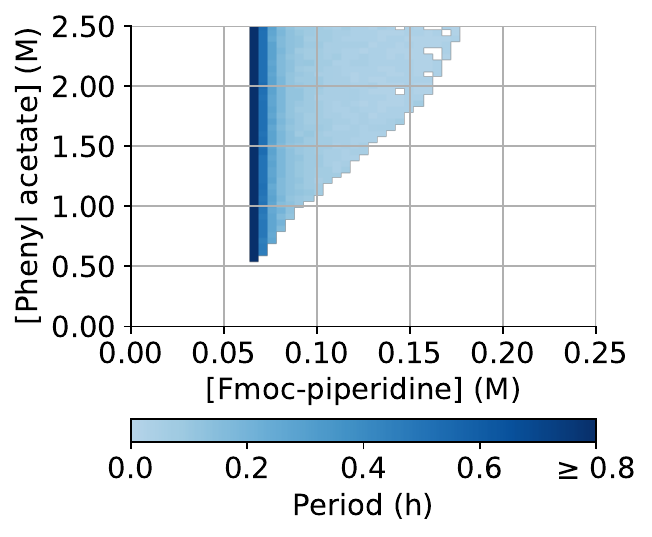}
         \caption{}
    \end{subfigure}
    \hfill
        \begin{subfigure}[b]{0.32\textwidth}
         \centering
         \includegraphics[width=\textwidth]{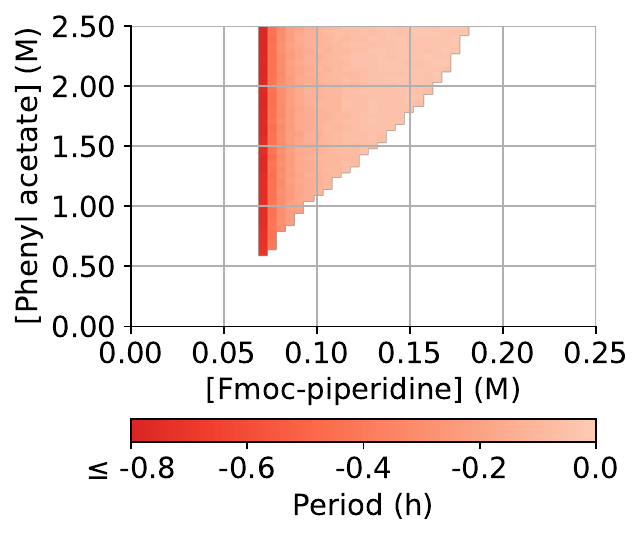}
         \caption{}
    \end{subfigure}
    \hfill
        \begin{subfigure}[b]{0.32\textwidth}
         \centering
         \includegraphics[width=\textwidth]{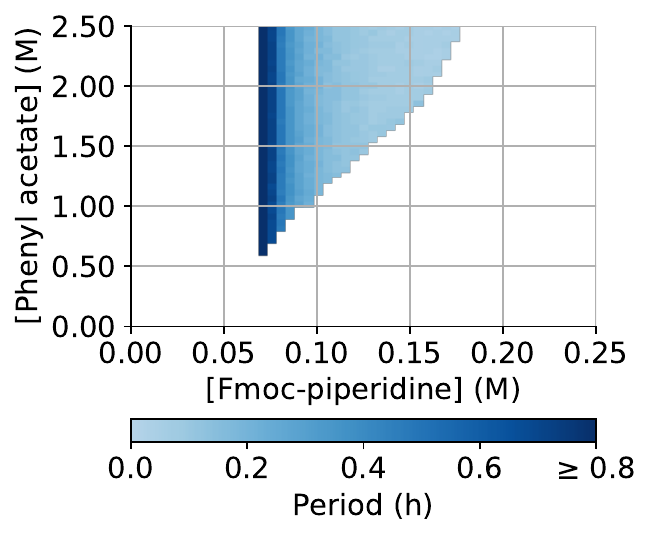}
         \caption{}
    \end{subfigure}
    \caption{The difference between the predicted oscillation spaces. (a), the predictions of the nODE trained on the full time series minus the theoretical predictions. (b), the predictions of the nODE trained on the full time series minus those of the nODE trained on half of the time series. (c) the predictions of the nODE trained on half of the time series minus the theoretical predictions. }
    \label{fig: phase diagram differences}
\end{figure}

While the nODEs are no better predictors of the type of oscillation than the theoretical model, they provide a better estimate of the oscillation period. As highlighted in Figure \ref{fig: phase diagram differences}, the predicted oscillation spaces differ in the size of the periods. In particular, the periods predicted by the nODE trained on the first half of the time series generally predict the largest periods, followed by the nODE trained on the full time series and the theoretical model. 

The trend found in Figure \ref{fig: phase diagram differences} recurs in Table \ref{tab: periods}: the nODE-based models tend to predict larger periods than the theoretical model (abbreviated with ODE). As a result, the periods predicted by the nODEs are a more accurate representation of the experimentally observed period than the predictions based on the theory alone.

\begin{table}[H]
    \centering
     \begin{adjustbox}{width=\textwidth}
    \begin{tabular}{c|c|c|c|c|c}
    \toprule
       \multicolumn{2}{c|}{Experimental setting} & Observed period  & ODE & nODE & nODE half \\
       \cline{1-2}
       $[$Fmoc-piperidine$]$ (M) & $[$Phenyl acetate$]$ (M) & (h) & (h) & (h) & (h)\\
       \hline
0.08 & 1.8 & 4.3 &3.52 &3.70 &4.00 \\
0.1 & 1.4 & 2.7 &2.37 &2.45 &2.55 \\
0.1 & 2.2 & 3.6 &2.33 &2.40 &2.52 \\
         \bottomrule
    \end{tabular}
    \end{adjustbox}
    \caption{The observed and predicted periods of different experimental settings. }
    \label{tab: periods}
\end{table}

\section{Conclusions}
The real-world behaviour of chemical species during a reaction can be modelled with a system of ODEs. The quality of this model is affected by the theoretical models that describe the chemical system, which can be incomplete. To account for the differences between the experimental data and the theoretical model, we combine dynamic systems modelling and deep learning in the form of neural ODEs. 

Our results demonstrate the applicability of a nODE for modelling aperiodic and oscillating concentrations. The separate neural network contributions provide an insight into the nature of the residuals between the theoretical model and the data. 

The neural network contributions may be hard to interpret, as they aggregate over all forms of noise within the system. Therefore, we focused on predicting the periodicity and phase space of an oscillating CRN by transferring the dynamics learned by the network between experiments. The nODEs do not outperform the theoretical ODE when classifying the regime of sustained oscillations. Despite training on experiments that represent various locations within the phase diagram, the neural network is unable to learn the dynamics of the transition between stable and decaying oscillations. We believe that additional experiments would be necessary in order to provide more training data that capture the deviations from ideal behaviour. Nonetheless, the nODEs are a valuable addition to the theoretical model, as they are a better predictor of the period of the oscillations. We advocate training the nODEs in conjunction with the development of the experimental systems and use the nODE predictions as a guide to which additional experiments would provide maximum improvement in nODE prediction. Altogether, we expect nODE predictions to help identify shortcomings of existing theoretical models and assist in the future design of functional CRNs.

\paragraph{Data and Software Availability}
The source code associated with the generation of synthetic data and the methods presented is available on GitHub: \url{https://github.com/KachmanLab/ChemicalReactionNetworks}. All software was prepared by the Radboud University authors. The experimental data used in this work have been publicly made available by ter Harmsel et al. and can be found online: \url{https://www.nature.com/articles/s41586-023-06310-2#Sec4}\cite{catalyticOscillator}.  

\paragraph{Supporting Information} Extra information on neural networks, chemical reaction networks and modelling dynamical systems; pseudocode of the algorithms presented in the methods; and the additional experimental results considering experimental data (PDF). 

\paragraph{Acknowledgements}
We acknowledge funding from the National Growth Fund project “Big Chemistry” (1420578), funded by the Ministry of Education, Culture and Science. In addition, Tal Kachman and Anna C.M. Thöni acknowledge funding from the European Laboratory for Learning and Intelligent Systems (ELLIS). 

\bibliography{main} 

\clearpage

\section{Supporting Information}
\section{Background}
\subsection{Artificial Neural Networks}
A neural network is a universal approximator that consists of the repeated application layers. Each layer combines a linear and nonlinear transformation $\mathbb{R}^D \to \mathbb{R}^M$. In a multilayer perceptron, a popular architecture, the output of the $j^{th}$ layer, where $j \in 1, ..., M$, is described by Equation \ref{eq: nn layer} \cite{bishop2006pattern}. \begin{equation}
    a_j = b\left(\sum^D_{i=1} w_{ji}x_i + w_{j0}\right)
    \label{eq: nn layer}
\end{equation}
where $w_{ji}$ represents the weight that is applied to the input $x_i$ and $w_{j0}$ is the bias. $b$ is a nonlinear, layer-specific activation function. The architecture of the network, i.e. the number of the layers, their widths $D$ and activation functions $b$, is determined prior to the training process. In contrast, the values of the weights $w$ are learned from the training data. 

\paragraph{Network Training} The data needed to train a neural network consists of a set of inputs $\{x_i\}$ with $i \in 1, ..., N$ and their corresponding labels $y_i$. The goal of the training process is to find the parameters $\theta$ that minimize the error function between the prediction of the network $f_\theta(x_i)$ and the labels $y_i$. In this work, we use the mean squared error (MSE) to calculate the loss of the network. 
\begin{equation}
    E = \frac{1}{N} \sum_{i=1}^N \left(f_\theta(x_i) - y_i\right)^2
\end{equation}
After each forward pass $f_\theta(x_i)$, the values of the weights are updated in the direction of the negative gradient that is associated with the error function. The magnitude of this update is regulated by the learning rate \cite{bishop2006pattern}.

\paragraph{Recurrent Neural Networks}
The information flow within the MLP described above is unidirectional: the outcomes $\bm{a}$ are only propagated to the subsequent layers. Alternatively, the information flow can be bidirectional, where the outcomes of a layer are added to subsequent inputs to the same layer. As such, the networks are provided with a sort of ``memory component". Within the class of these recurrent neural networks, the long short-term memory (LSTM) is a popular architecture \cite{hochreiter1997long}.
\subsection{Chemical Reaction Networks}
A chemical reaction network (CRN) is a framework that describes the relationships between chemical reactions \cite{Angeli2009}. To illustrate, consider the theoretical reaction network $A \to B \to C$, consisting of the reactions $\mathcal{R} = \{R_1, R_2\}$:
\label{eq: bottleneck}
\begin{align}
\begin{split}
    R_1{:}\;\;A &\xrightarrow[]{k_1} B  \\
    R_2{:}\;\;B &\xrightarrow[]{k_2} C
\end{split}
\end{align}
In this process, a single unit of compound $A$ is converted to compound $B$ with a reaction rate $k_1$. Next, compound $B$ goes to compound $C$ with rate $k_2$. The three different compounds that take part in the reaction are referred to as the species, $\mathcal{S} = \{A, \; B, \; C\}$ with $n_s = |\mathcal{S}| =3$. In chemical reactions, the species on the left-hand side of the arrow are considered the reactants, whereas the species on the right-hand side are considered the products. 
\paragraph{Dynamical Modelling and mass action Kinetics}
Using a CRN and the relations between species that it describes, one can construct a system of ODEs to determine the rate of concentration change of the compounds involved. More specifically, if the mixture of compounds is well-stirred and the number of compounds is sufficiently high, it can be assumed that the reactions $\mathcal{R}$  occur according to an exponential probability distribution \cite{UpadhyayChemicalKinetics, Angeli2009}. As such, the rate $r_i$ of each reaction $R_i \in \mathcal{R}$ can be described according to the mass action kinetics:
\begin{equation}
    r_i(S) = k_i \prod_{j=1}^{n_s} [S_j]^{a_{ij}}
    \label{eq: mass_action_kinetics}
\end{equation}
In Equation \ref{eq: mass_action_kinetics}, $k_i$ is the reaction rate coefficient,\footnote{In this work, the reaction rate coefficient is regarded as a constant as it is independent of the time $t$ \cite{levine2009molecular}, the main control variable used.} $S$ is the vector of all species, $[S_j]$ is the concentration of species $S_j$ and $a_{ij}$ is the stoichiometric coefficient of species $S_j$ in reaction $R_i$. The rate coefficient $k_i$ can be determined by experiment or simulation \cite{UpadhyayChemicalKinetics}.\\[3mm]
Now, let $C_i$ be the reactants and $C_i'$ the products of $R_i$. Using this definition, we can determine the state change of $R_i$: if reaction $R_i$ can be described by $C_i \to C_i'$, the state change can be summarized by $a_i \to a_i'$. Here, $a_i$ is the vector of the stoichiometry coefficients of the set of reactants or products from reaction $R_i$. $a_i$ is grouped by the species. For example, for $R_1$ it holds that $a_1 = [1, 0, 0]^T$ and $a_1' = [0, 1, 0]^T$. Using the state change, the rate of the concentration change that is brought about by $R_i$ is defined as the state change times the rate $r_i$, which results in a vector of $n_s$ derivatives (Equation \ref{eq: change in concentration}). 

\begin{equation}
    \frac{d[S]}{dt} = (a_i' - a_i) \times k_i \prod_{j=1}^{n_s} [S_j]^{a_{ij}}
    \label{eq: change in concentration}
\end{equation}
For CRNs with multiple reactions, the rate of change in the total concentration of a species is provided by the sum over all reactions $R_i \in \mathcal{R}$. Thus, returning to the CRN from Equation \ref{eq: bottleneck} a final time, if the reaction rates are assumed to be $k_1$ and $k_2$, one can describe the rate of the change in concentration of the three species as in Equation \ref{eq: vector_field}.
    \begin{align}
    \begin{split}
        \label{eq: vector_field}
        \frac{d[A]}{dt} &= -k_1[A]\\
        \frac{d[B]}{dt} &= k_1[A] - k_2[B]\\
        \frac{d[C]}{dt} &= k_2[B]
        \end{split}
\end{align}
\noindent
The vector field described by Equation \ref{eq: vector_field} can, in combination with the initial concentrations of each species, be used to obtain an estimate of the concentrations over time by solving the initial value problem (IVP).
\subsection{Modelling Dynamical Systems}
\paragraph{Expressing the Initial Value Problem in Neural Vector Fields.}
To provide an introduction to nODEs, we first define the IVP in more formal terms. Let $y{:}\; [0, T] \to \mathbb{R}^m$ be the unknown solution described by the initial point $y_0 \in \mathbb{R}^m$ and differential equation $h(t, y(t)){:}\; \mathbb{R} \times \mathbb{R}^m \to \mathbb{R}^m$. Using these terms, the IVP is formalized in Equation \ref{eq: IVP_appendix}.
\begin{equation}
    y(0) = y_0 \;\;\;\;\;\;\;\; \frac{dy}{dt}(t) = h(t, y(t))
    \label{eq: IVP_appendix}
\end{equation}
While the solution to Equation \ref{eq: IVP} equals $y = \int_0^T h(t, y(t))$, systems of differential equations are often solved using numerical methods due to the unavailability of the antiderivative of $h$ \cite{Calculus2013}. The simplest method to solve an IVP is to discretize the derivative according to the Euler discretization \cite{lu2018beyond}.
\begin{equation}
    \frac{dy}{dt}(t_n) \approx \frac{y(t_{n+1}) - y(t_n)}{\Delta t}
\end{equation}
\noindent
Using the discretization above, the vector field from Equation \ref{eq: IVP_appendix} can be rewritten to 
\begin{equation}
    y(t_{n+1}) = y(t_n) + \Delta t \cdot h(t, y(t_n))
    \label{eq: discretized IVP}
\end{equation}
\noindent
If $h$ is represented by the neural network $f_\theta$, then $\Delta t$ can be absorbed into the neural network term to yield Equation \ref{eq: discrete_resnet} \cite{kidger2021}. 
\begin{equation}
    y(t_{n+1}) = y(t_n) + f_\theta(t_n, y(t_n))
    \label{eq: discrete_resnet}
\end{equation}

\citeauthor{chen2018neural}\ point out that Equation \ref{eq: discrete_resnet} bears similarities with the formulation of a residual neural network \cite{he2015deep}, where the addition of $ y(t_n)$ can be interpreted as the skip connections within the residual block. Fuelled by this observation, \citeauthor{chen2018neural} introduce neural ODEs, which they consider a residual network with a ``continuous depth" \cite{chen2018neural}. 

\paragraph{Universal Differential Equations}
\citeauthor{rackauckas2020universal} extend the nODEs by incorporating domain knowledge into the model. Their proposed universal differential equation combines the fully data-driven with a theoretical vector field $h$ that is specified beforehand \cite{rackauckas2020universal}.
\begin{equation}
    \frac{dy}{dt}(t_{n+1}) = h_\kappa(t, y(t_n)) + f_\theta(t, y(t_n))
    \label{eq: UDE_appendix}
\end{equation}
In Equation \ref{eq: UDE_appendix}, $\kappa$ represents the paremeters of $h$. When vector fields based on CRNs are considered, then $h$ represents the vector field based on the mass action dynamics of the CRN. In that case, $\kappa$ denotes the vector of reaction rate coefficients $\bm{k}$.

\paragraph{Differential Equation Solvers and Stiffness} Solving a system of differential equations in a stepwise fashion as described above introduces a limitation: the performance of the ODE solvers may suffer from the stiffness of the IVP at hand. That is, IVPs that contain regions of strongly increasing or decreasing dynamics, an often-occurring phenomenon within chemical kinetics, can cause numerical instability \citeauthor{CHEMSODE}. Whereas \citet{ji2021autonomous} used soft constraints in their physics-informed neural network to alleviate this instability, it can be countered within nODEs by using an implicit ODE solver \cite{DIRK}.  
\clearpage
\section{Pseudocode}
\begin{minipage}{\linewidth}
\begin{algorithm}[H]
\caption{Neural ODE training}
\label{alg:training process}
\raggedright
\textbf{Input} \\
\hspace*{3mm}$N_{epochs}$: the number of epochs\\
\hspace*{3mm}$T$: the time of the last measurement\\
\hspace*{3mm}$\mathcal{D}$: the dataset $\{N \times (t_0,\, \bm{y}_0),\, (t_1,\, \bm{y}_1),\, \dots, \, (T,\, \bm{y}_T)\}$\\
\hspace*{3mm}$f_\theta$: the neural network with parameters $\theta$\\
\hspace*{3mm}$h_{\bm{\kappa}}$: the theoretical system of ODEs\\
\textbf{Output} \\
\hspace*{3mm}The trained model $f_\theta$
 \vspace{1mm}
\begin{algorithmic}
\FOR{epoch in $N_{epochs}$}
\FOR{$\bm{x}, \bm{y}$ in $\mathcal{D}$}
\STATE $t_{s} \gets t_0$
\WHILE{$t_s < T$} 
\IF{$t_s$ is $t_0$}
\STATE $\hat{\bm{y}}_{t_{s}} \gets \bm{y}_0$ 
\ENDIF
\STATE $\Delta t \gets \text{ODEsolver}(\hat{\bm{y}}_{t_{s}}, t_{s})$ 
\STATE $\hat{\bm{y}}_{t_{s} + \Delta t} \gets h_{\bm{\kappa}}(\hat{\bm{y}}_{t_{s}}, \; \Delta t) +  f_\theta(\hat{\bm{y}}_{t_{s}},\; \Delta t)$
\STATE $t_{s} \gets t_{s} + \Delta t $
\ENDWHILE
\STATE $\mathcal{L} = \text{MSE}\left(\hat{\bm{y}}_{t_0:T}, \bm{y}\right)$
\STATE Update $\theta$ by taking a gradient descent step on  $\mathcal{L}$
\ENDFOR
\ENDFOR 
\end{algorithmic}
\end{algorithm}
\end{minipage}

\noindent
\begin{minipage}{\linewidth}
\begin{algorithm}[H]
\caption{Neural network contribution}
\label{alg: nn_contribution}
\textbf{Input} \\
\raggedright
\hspace*{3mm}$\hat{\bm{y}}$: the concentrations predicted by the neural ODE.\\
\hspace*{3mm}$f_\theta$: the neural network with parameters $\theta$\\
\textbf{Output} \\
\hspace*{3mm}$c_{f_\theta}$: the neural network contribution
 \vspace{1mm}
\begin{algorithmic}
\FOR{$\hat{\bm{y}}_t$ in $\hat{\bm{y}}$}
\STATE $c_{f_\theta,t} \gets f_\theta(\hat{\bm{y}}_t)$
\ENDFOR 
\end{algorithmic}
\end{algorithm}
\end{minipage}
\clearpage
\section{Supplementary Figures}\label{Appendix: experimental}
This section presents all results that have been found when considering the experimental data, summarizing the standard fit, open system and missing reaction experiments with the single-pulse and oscillating data. The section serves as a reference without providing an in-depth description of all figures. Instead, all conclusions drawn from these results are provided in the main document. 
\subsection{Single-pulse Experiment}\label{appendix: single-pulse}
\paragraph{Standard Fit}\phantom{text}
\begin{figure}[H]
\centering
\begin{subfigure}[b]{0.48\textwidth}
\includegraphics[width=\textwidth]{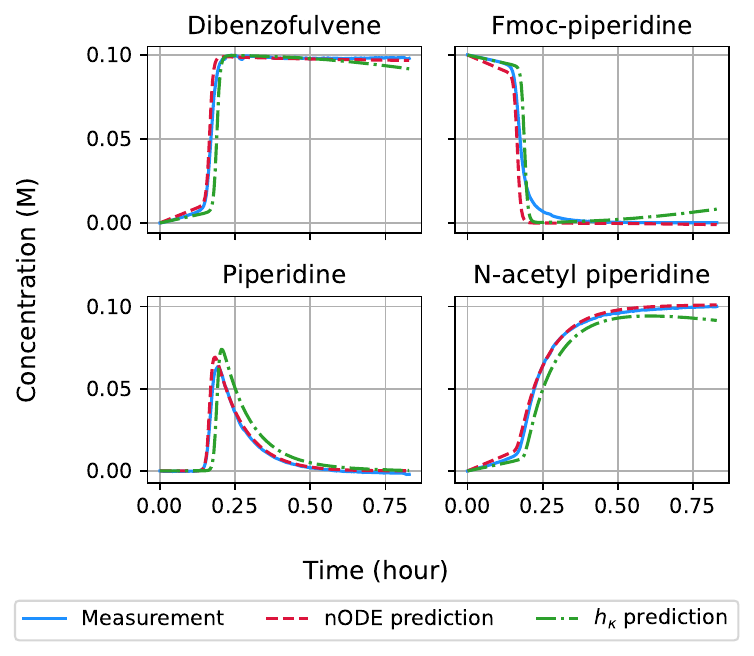}
\caption{} \end{subfigure}
\hfill
\begin{subfigure}[b]{0.48\textwidth}
    \includegraphics[width=\textwidth]{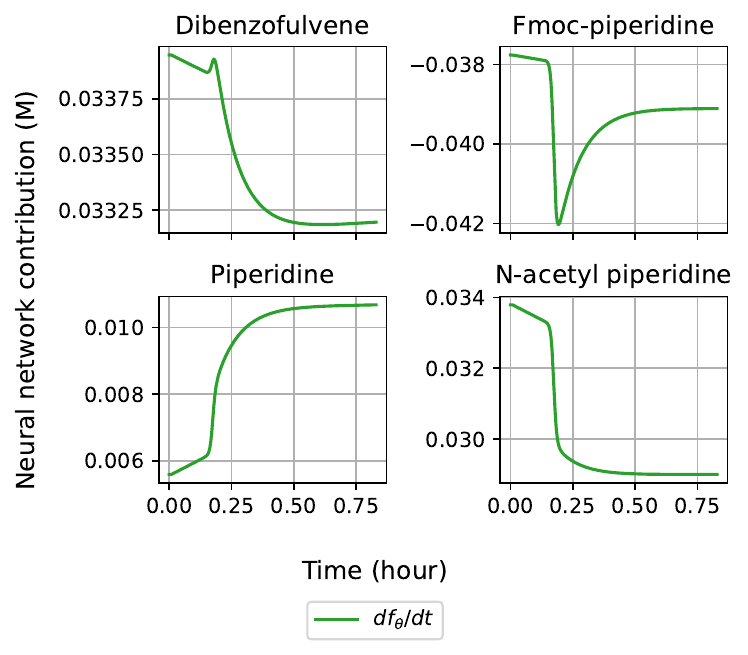}
\caption{} \end{subfigure}
\vspace{-3mm}
\caption[The predictive performance of the nODE on the single-pulse data]{The predictive performance of the nODE on the single-pulse data. (a), the experimental measurements, predictions (solid blue) from the nODE (dashed red) and $h_\kappa$ (dashdotted green). The four panes illustrate the molar concentrations for the species dibenzofulvene, Fmoc-piperidine, piperidine and N-acetyl piperidine. (b), the neural network contribution for the four measured species.}
\end{figure}
\clearpage
\paragraph{Open System}\phantom{text}
\begin{figure}[H]
\centering
\begin{subfigure}[b]{0.48\textwidth}
\includegraphics[width=\textwidth]{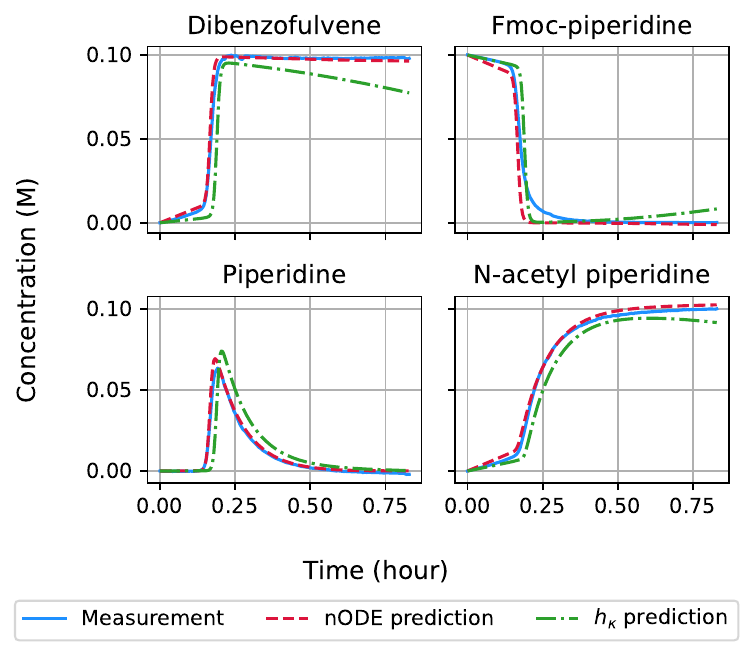}
\caption{} \end{subfigure}
\hfill
\begin{subfigure}[b]{0.48\textwidth}
\includegraphics[width=\textwidth]{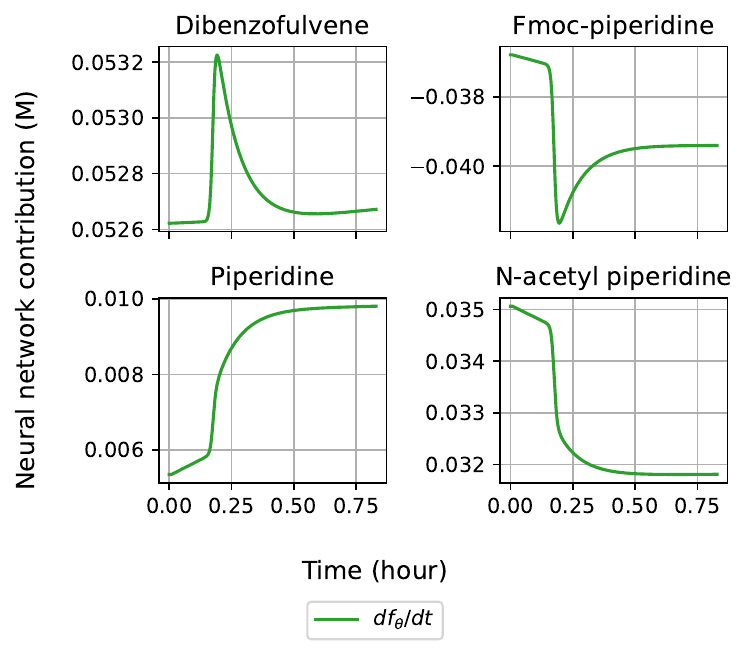}
\caption{} \end{subfigure}
\vspace{-3mm}
\caption[The predictive performance of the nODE on the single-pulse data in the open system]{The predictive performance of the nODE on the single-pulse data in the open system. (a), the experimental measurements, predictions (solid blue) from the nODE (dashed red) and $h_\kappa$ (dashdotted green). The four panes illustrate the molar concentrations for the species dibenzofulvene, Fmoc-piperidine, piperidine and N-acetyl piperidine. (b), the neural network contribution for the four measured species.}
\end{figure}
\clearpage
\paragraph{Missing Reactions}\phantom{text}
\begin{figure}[H]
\centering
\begin{subfigure}[b]{0.48\textwidth}\includegraphics[width=\textwidth]{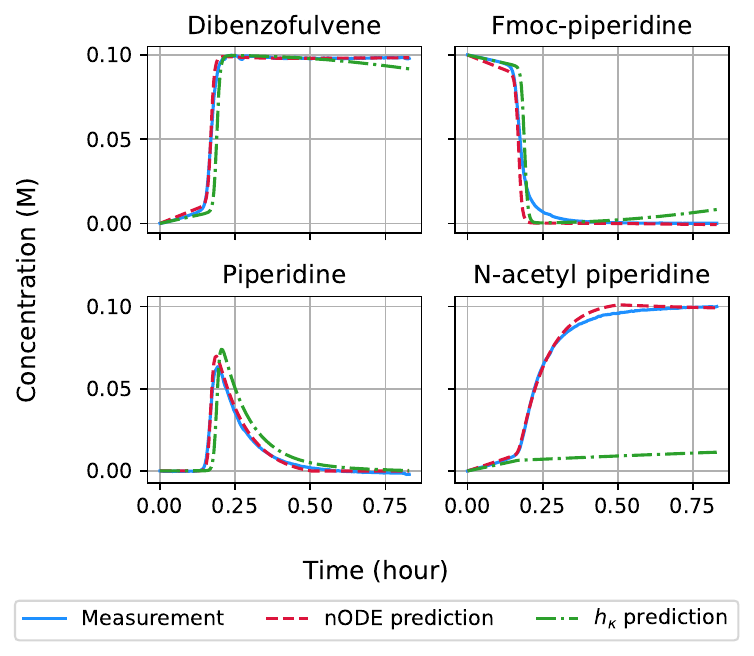}
\caption{} \end{subfigure}
\hfill
\begin{subfigure}[b]{0.48\textwidth}
\includegraphics[width=\textwidth]{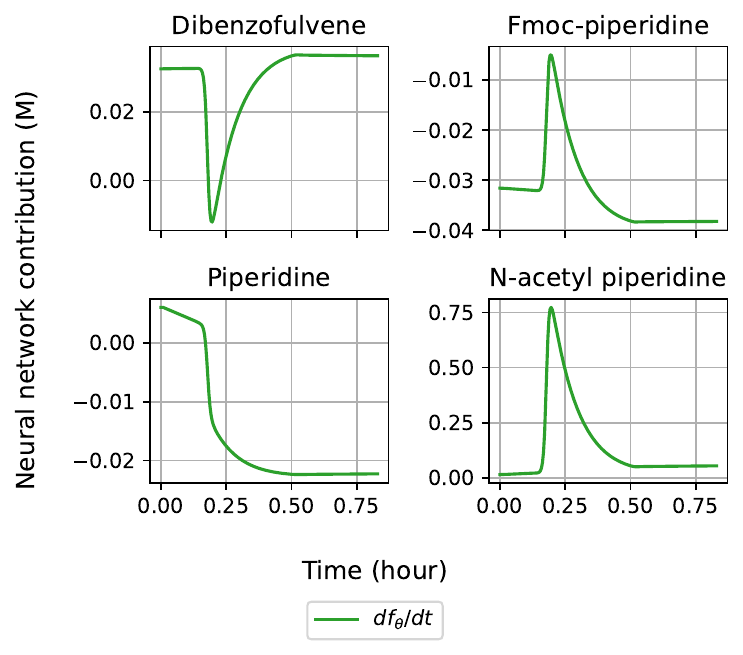}
\caption{} \end{subfigure}
\vspace{-3mm}
\caption[The predictive performance of the nODE with an incomplete vector field on the single-pulse data in the open system]{The predictive performance of the nODE with an incomplete vector field on the single-pulse data in the open system. (a), the experimental measurements, predictions (solid blue) from the nODE (dashed red) and $h_\kappa$ (dashdotted green). The four panes illustrate the molar concentrations for the species dibenzofulvene, Fmoc-piperidine, piperidine and N-acetyl piperidine. (b), the neural network contribution for the four measured species.}
\end{figure}
\clearpage
\subsection{Sustained Oscillations}\label{appendix: oscillating}
\paragraph{Standard Fit}\phantom{text}
\begin{figure}[H]
\centering
\begin{subfigure}[b]{0.67\textwidth}\includegraphics[width=\textwidth]{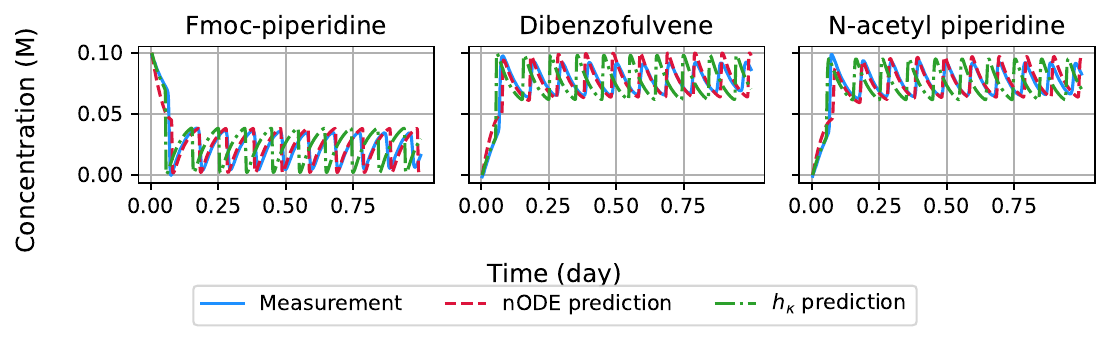}
\caption{} \end{subfigure}
\hfill
\begin{subfigure}[b]{0.67\textwidth}\includegraphics[width=\textwidth]{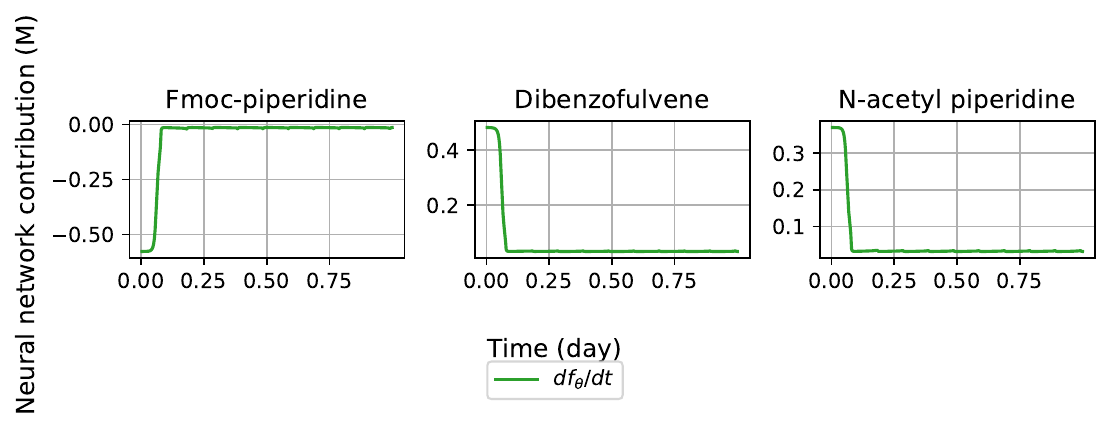}
\caption{} \end{subfigure}
\vspace{-3mm}
\caption[The predictive performance of the nODE on the oscillating data]{The predictive performance of the nODE on the oscillating data. (a), the experimental measurements, predictions (solid blue) from the nODE (dashed red) and $h_\kappa$ (dashdotted green). The three panes illustrate the molar concentrations for the species Fmoc-piperidine, dibenzofulvene and N-acetyl piperidine. (b), the neural network contribution for the four measured species.}
\end{figure}
\clearpage
\paragraph{Open System}\phantom{text}
\begin{figure}[H]
\centering
\begin{subfigure}[b]{0.67\textwidth}\includegraphics[width=\textwidth]{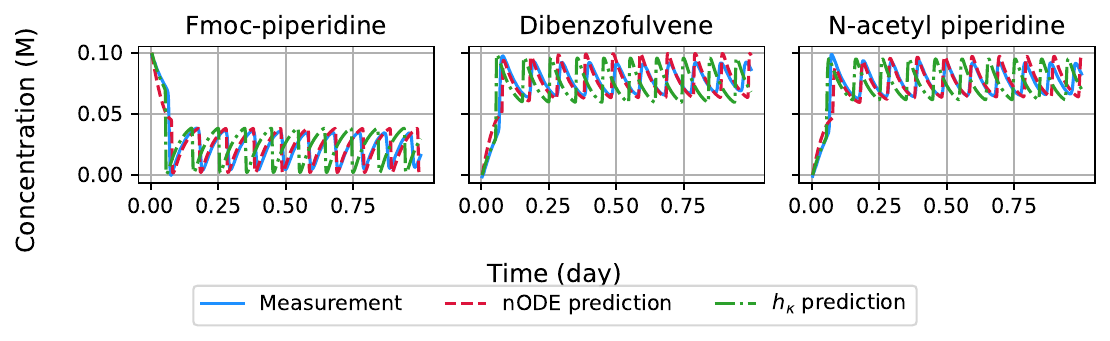}
\caption{} \end{subfigure}

\begin{subfigure}[b]{0.67\textwidth}\includegraphics[width=\textwidth]{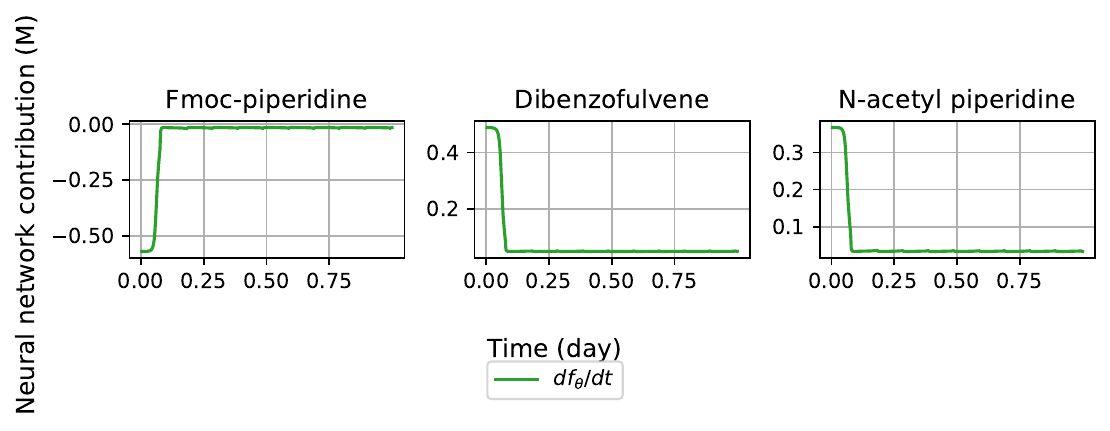}
\caption{} \end{subfigure}
\vspace{-3mm}
\caption[The predictive performance of the nODE on the oscillating data]{The predictive performance of the nODE on the oscillating data in the open system. (a), the experimental measurements, predictions (solid blue) from the nODE (dashed red) and $h_\kappa$ (dashdotted green). The three panes illustrate the molar concentrations for the species Fmoc-piperidine, dibenzofulvene and N-acetyl piperidine. (b), the neural network contribution for the four measured species.}
\end{figure}
\clearpage
\paragraph{Missing Reactions}\phantom{text}
\begin{figure}[H]
\centering
\begin{subfigure}[b]{0.67\textwidth}\includegraphics[width=\textwidth]{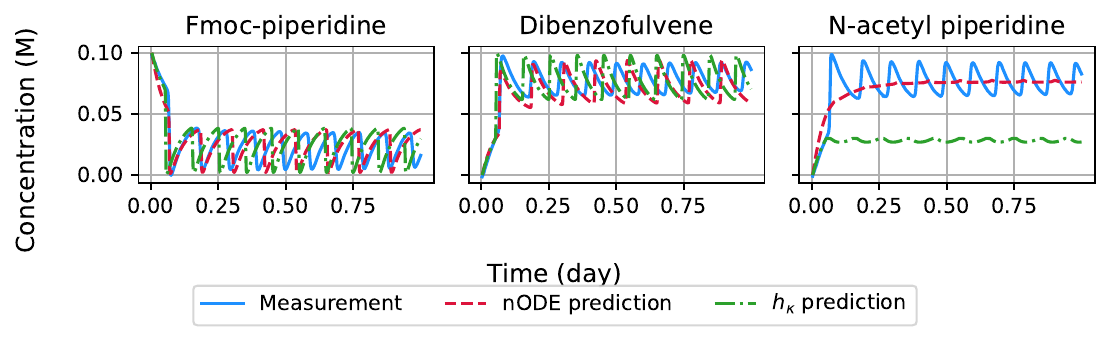}
\caption{} \end{subfigure}

\begin{subfigure}[b]{0.67\textwidth}\includegraphics[width=\textwidth]{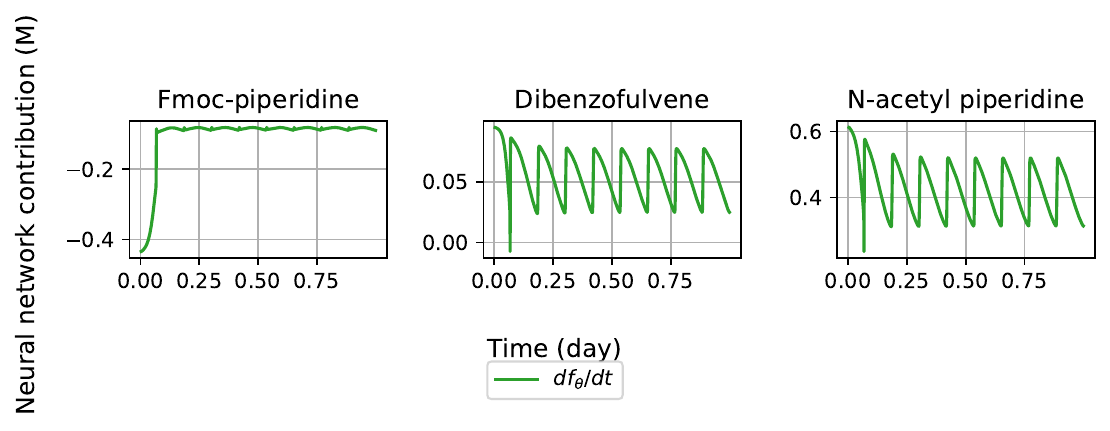}
\caption{} \end{subfigure}
\vspace{-3mm}
\caption[The predictive performance of the nODE on the oscillating data]{The predictive performance of the nODE on the oscillating data of the missing reactions dataset. (a), the experimental measurements, predictions (solid blue) from the nODE (dashed red) and $h_\kappa$ (dashdotted green). The three panes illustrate the molar concentrations for the species Fmoc-piperidine, dibenzofulvene and N-acetyl piperidine. (b), the neural network contribution for the four measured species.}
\end{figure}
\clearpage

\end{document}